\documentclass[aps, showpacs,pra,twocolumn] {revtex4}
\usepackage{hyperref}
\usepackage{amsmath }
\usepackage{amssymb}
\usepackage{epsfig}
\usepackage{natbib}
\newcommand*{\be}{\begin{equation}}
\newcommand*{\ee}{\end{equation}}

\begin{document}
\bibliographystyle{revtex}

\title{ New approximate radial wave functions
 \\
 for power-law potentials }

\author{Vladimir Kudryashov }
 \email{ kudryash@dragon.bas-net.by}
\affiliation{B.I. Stepanov Institute of Physics, National Academy
of Sciences of Belarus,\\  68 Nezavisimosti Ave., 220072 Minsk,
Belarus}

\received{ 25 May  2006 }

\begin{abstract}Radial wave functions for power-law potentials are
approximated with the help of power-law substitution and explicit
summation of the leading constituent WKB series. Our approach
reproduces the correct behavior of the wave functions at the
origin, at the turning points and far away from  the turning
points
\end{abstract}


\maketitle

\section{Introduction}

Many interesting physical problems require solving the
Schr\"odinger equation for spherically symmetric potentials
$V(r)$.
 The separated radial Schr\"odinger equation can be
written in the form
\begin{equation}
-\frac{{\hbar^2}}{2 m} \frac{d^2 \psi(r)}{d r^2} +  \left( V(r) +
V_c(r) - E\right)\psi(r) = 0
\end{equation}
which is identical to the one-dimensional Schr\"odinger equation
with an effective potential given by the sum of the origin
potential $V(r)$ and the centrifugal potential
\begin{equation}
V_c(r) = \frac{\hbar^2 l (l+1)}{2 m r^2} .
\end{equation}
In the present work we consider power-law potentials
\begin{equation}
V(r) =  \alpha r^k ,\quad  \alpha > 0 , \quad k \geq 1
\end{equation}
which are very important in particle physics. It is known that
only several potentials permit the exact solutions of the
Schr\"odinger equation.  Usually realistic physical calculations
are impossible without different approximation techniques. One of
the earliest and simplest methods of obtaining
 approximate eigenvalues and eigenfunctions of the radial Schr\"odinger
  equation is the WKB method (see, e.g., \cite{lang,krie} and references therein).

 It is known \cite{lang,from} that a suitable transformation of the initial
equation improves  results of an approximation technique. We
examine the power-law substitutions
\begin{equation}
r = q^s , \quad s>0 , \quad \psi(r) =r^{(s - 1)/{2 s}} \Psi(q) .
\end{equation}
The transformed equation is
\begin{equation}
-{\hbar^2} \frac{d^2 \Psi(q)}{d q^2} + Q(q) \Psi(q) =0
 \end{equation}
 where
\begin{eqnarray}
 Q(q) =\frac{2 m s^2}{q^2} \left(\alpha q^{(2+k) s}
 - E q^{2 s}\right) \nonumber \\
  + \frac{\hbar^2}{q^2}\left(s^2(l + 1/2)^2 - \frac{1}{4}\right) .
\end{eqnarray}
Of course, the exact solutions do not depend on some substitution.
However, we are interesting in the approximate solutions. R. E.
Langer \cite{lang} has used the particular case $ s=2 $ when he
was applying the WKB approximation to the Kepler problem.

 The WKB approach deals with the
logarithmic derivative
\begin{equation}
 Y(q) =\frac{d \ln \Psi(q)}{d q}
\end{equation}
that satisfies  the nonlinear Riccati equation
\begin{equation}
-\hbar^2\left( \frac{dY(q)}{dq} + Y^2(q)\right) + Q(q) =0
\end{equation}
where $Q(q)$ is an arbitrary function of $q$ (naturally not only
the special expression (6)). The WKB series
\begin{equation}
 Y_{as}^{\pm}(q) = \hbar^{-1}\left(
\pm Q^{1/2} +
 \sum_{n = 1}^{\infty} \hbar^n Y^{\pm}_n(q)\right)
\end{equation}
are the asymptotic expansions in powers of Plank's constant
$\hbar$ of two independent particular solutions of  the Riccati
equation.  The usual WKB approximations
 $$
 Y_{WKB}^{\pm}(q) =
  \hbar^{-1}\left(\pm   Q^{1/2} +
 \sum_{n = 1}^{N}  \hbar^n Y^{\pm}_n(q)\right)
$$
 contain a finite number of leading terms $Y_n^{\pm}(q)$ from
the complete expansions $ Y^{\pm}_{as}(q)$. These approximations
are not valid at  the turning points where $Q(q) = 0$ and at the
origin $q =0$. While in most cases of improvements of the WKB
method (see, e.g., \cite{krie,from,seet,laks}) the main purpose is
to achieve highest accuracy in eigenvalue calculation for the
radial Schr\"odinger equation, our goal is to construct
satisfactory approximate eigenfunctions with the correct behavior
at the origin, at the turning points and far away from  the
turning points.

\section{New approximations for logarithmic derivatives}
The analysis of the well-known structure of the leading $Y_n^{\pm}(q)$
and recursion relations  \cite{bend1,bend2} allows us to
reconstruct the asymptotic WKB series as the infinite sums
\begin{equation}
 Y^{\pm}_{as}(q) = \pm  \hbar^{-1} Q^{1/2} + \sum^{\infty}_{j =
1}Z^{\pm}_{as,j} (q)
\end{equation}
of new constituent (partial) asymptotic series $
Z^{\pm}_{as,j}(q)$ \cite{kudr1,kudr2}.

The complete series $Y^{\pm}_{as}(q)$ are approximated by a finite
number of leading constituent series $Z^{\pm}_{as,j}(q)$ in
contrast to the use of a finite number of leading  terms
$Y^{\pm}_n(q)$ in the conventional WKB approach. Using notation
\begin{equation}
a(q) = \frac{1}{\hbar^{2/3}} \frac{Q(q)}{|d Q(q)/dq|^{2/3}},
\end{equation}
\begin{equation}
b_1(q) = \frac{1}{\hbar^{2/3}} \frac{d Q(q)/dq}{|d Q(q)/dq|^{2/3}}
,
\end{equation}
\begin{equation}
 b_2(q) = \frac{d^2Q(q)/dq^2}{d Q(q)/dq}
\end{equation}
we are able to rewrite two first leading constituent series in the
form
\begin{eqnarray}
 \pm \hbar^{-1} Q^{1/2} + Z^{\pm}_{as,1}(q) +
Z^{\pm}_{as,2}(q) \nonumber \\
 =   b_1(q)  y^{\pm}_{as,1}(a) +
 b_2(q) y^{\pm}_{as,2}(a).
\end{eqnarray}

Direct verification shows that the series $y^{\pm}_{as,1}(a)$ and
$y^{\pm}_{as,2}(a)$ satisfy  equations
\begin{equation}
 \frac{dy^{\pm}_{as,1}}{da} + (y^{\pm}_{as,1})^2 = a ,
\end{equation}
\begin{equation}
 \frac{dy^{\pm}_{as,2}}{da} + 2y^{\pm}_{as,1}y^{\pm}_{as,2}  =
 \frac{1}{3}\left( 2a\frac{dy^{\pm}_{as,1}}{da} -
y^{\pm}_{as,1}\right) .
\end{equation}
Eq.(15) is the Riccati equation for the logarithmic derivatives of
linear combinations of the well-studied Airy functions ${\rm
Ai}(a)$ and ${\rm Bi}(a)$ \cite{abr} .

The particular expressions
 $$
 y^{\pm}_1(a) = \frac{d}{da} \ln\left({\rm Ai}(a) \pm i{\rm
 Bi}(a)\right),
$$
$$
  y^-_1(a) = \frac{d}{da} \ln {\rm Ai}(a) , \quad
y^+_1(a) = \frac{d}{da} \ln {\rm Bi}(a)
$$
correspond to the conventional WKB series .

Generalization consists in the use of formulas
 \begin{equation}
 y_1(a;t) = \frac{d}{da}\ln\left({\rm Ai}(a) + t {\rm
Bi}(a)\right) ,
\end{equation}
 \begin{equation}
  y_2(a;t) =
\frac{1}{30} \left[-8 a^2 (y_1(a;t))^2 -3  
-4 a y_1(a;t) + 8 a^3  \right]
\end{equation}
with a mixture parameter $t$. As a result we get  new
approximate logarithmic derivative
\begin{equation}
Y_{app}(q) = Y(q;t) 
 = b_1(q)  y_1(a;t) +
 b_2(q) y_2(a;t)
\end{equation}
which satisfies the following equation
\begin{widetext}
\begin{eqnarray}
-\hbar^2\left( \frac{dY(q;t)}{dq} + Y^2(q;t)\right) + Q(q) =
  -\hbar^2\left[\left(\frac{d^3Q(q)/dq^3}{dQ(q)/dq}\right)y_2(a;t) \right.
 \nonumber \\
\left. +
\left(\frac{d^2Q(q)/dq^2}{dQ(q)/dq}\right)^2\left(y^2_2(a;t)
-\frac{8}{3} y_2(a;t) + \frac{4}{3} a y_1(a;t) y_2(a;t) -
\frac{1}{6}\right) \right]
\end{eqnarray}
\end{widetext}
 instead of Eq. (8).

 It is not surprising that the asymptotics of our approximation
coincide with the WKB asymptotics far away from  the turning
points. At the same time our approximation reproduces the known
\cite{bend2} satisfactory approximation near  the turning points.

\section{New approximations for wave functions
}
Now we can construct the approximate radial wave functions for the
bound states in the case of a power-law potential when $Q(q)$ is
of the form (6).

 First, we must reproduce the correct limiting
behavior at the origin. In this case we have the following exact
expressions $(r \rightarrow 0, q \rightarrow 0)$
 $$
 \psi_{ex}(r) \rightarrow r^{l+1} , \quad
\Psi_{ex}(q) \rightarrow q^{s l + (s+1)/2},
 $$
 \begin{equation}
  Y_{ex}(q)
\rightarrow \frac{s l + (s+1)/2}{q} .
\end{equation}
At the same time we can derive relations
 $$
  a(q) \rightarrow a_0 =
\left(\frac{s^2}{4}(l+1/2)^2 -\frac{1}{16}\right)^{1/3} ,
 $$
 $$
 b_1(q) \rightarrow -\frac{2 a_0}{q} , \quad b_2(q) \rightarrow
-\frac{3}{q} ,
 $$
\begin{equation}
 Y(q;t) \rightarrow  -\frac{1}{q}\left(2 a_0 y_1(a_0;t) +3
y_2(a_0;t)\right)
\end{equation}
in the framework of our approach. We obtain the algebraic equation
for determining the value of $t$. Its solution is
\begin{equation}
 t_0 = \frac{-c(l,s) {\rm Ai}(a_0) + a_0 (d {\rm
Ai}(a_0)/d a_0)} {c(l,s) {\rm Bi}(a_0) - a_0( d {\rm Bi}(a_0)/d
a_0)}
\end{equation}
where $$ c(l,s) = 1 - \sqrt{1 + \frac{5}{4} \left(\frac{8 a^3_0 -
3}{10} + s(l +1/2) +1/2\right)} . $$

Two real turning points $q_-$ and $q_+$ ($Q(q_{\pm}) =0$) separate
three regions.

In the first region  where $0<q<q_- $ we select the unique
approximate particular logarithmic derivative $ Y(q;t_0)$. In the
second region where $q_- < q < q_+$ we must describe the
oscillatory solution of the original Schr\"odinger equation (1).
Therefore in this case we select two approximate particular
logarithmic derivatives $Y(q;+i)$ and $Y(q;-i)$. In the third
region where $q> q_+$ we must describe only the decreasing
solution of the original Schr\"odinger equation (1). Therefore in
this case we select the unique approximate particular logarithmic
derivative $Y(q;0)$. Note that in the case $l=0, s=1$ we put $q_-
= 0$.

Since  the turning points are ordinary nonsingular
points in our approach, no question of connection formulas arises in contrast
with the conventional WKB method. Matching particular solutions at
the turning points we obtain the continuous approximate radial
wave function
\begin{equation}
\psi_{app}(r) =  N_{app} r^{(s-1)/2s}  \Psi_{app}(q)
\end{equation}
 where $\Psi_{app}(q)$  is
represented by the following formulas
\begin{equation}
 \Psi_1(q) =  \cos{\phi}
 \exp\left( -\int_q^{q_-} Y(q';t_0)\, dq' \right)
\end{equation}
 if $0<q<q_- $ ,
 \begin{widetext}
\begin{eqnarray}
 \Psi_2(q) =  \exp \left( \int _{q_-}^q
 \frac{Y(q';+i) + Y(q';-i)}{2}\, d q' \right)
 \cos \left(\int_{q_-}^q \epsilon
 \frac{Y(q';+i) - Y(q';-i)}{2 i}\, d q' - \phi \right)
 \end{eqnarray}
 if $q_- < q < q_+$  ,
\begin{eqnarray}
 \Psi_3(q) = \frac{1}{2}(-1)^n
 \exp\left( \int_{q_+}^q Y(q';0)\, d q' \right)
 \exp\left( \int_{q_-}^{q_+} \frac{Y(q';+i) +
 Y(q';-i)}{2}\, d q'\right)
 \end{eqnarray}
 if $q> q_+$ .
 \end{widetext}
Here $q = r^{1/s},  \phi = \frac{\pi}{3} -  \arctan t_0 $,  $
\epsilon = \left( d Q(q)/dq\right)|d Q(q)/dq|^{-1}
 $ and
$N_{app}$ is a normalization constant.

We have the new quantization condition
 \begin{eqnarray}
 \int_{q_-}^{q_+}
\epsilon \frac{Y(q;+i) -
 Y(q;-i)}{2 i}\, d q \nonumber \\
= \pi (n + \frac{1}{3}) + \phi, \quad n = 0,1,2...
\end{eqnarray}
which determines the spectral value of $E$ implicitly.

Note that up to now a value of $s$ is not fixed. Numerical
experiment for the power-law potentials  ($k \geq 1$)  shows that
the best choice for $l=0$ is $s=1$ and the satisfactory common
choice for all $l >0$ is $ s= 2$.  Thus the approximate
eigenfunctions are determined completely and we can perform
verification.

\section{Application to power-law potentials}

 It is convenient to test our approximation with introducing the dimensionless quantities
$$
 x = \left(\frac{2 m \alpha}{\hbar^2}\right)^{1/(k+2)} r ,
$$
 $$
 e = \left(\frac{2 m}{\hbar^2 \alpha^{2/k}}\right)^{k(/k+2)} E .
$$
 Then the Schr\"odinger equation is rewritten in the form
$$
 \hat H  \psi(x) - e \psi(x) = 0
$$
 with the Hamiltonian
\begin{equation}
 \hat H = -\frac{d^2}{d x^2} + x^k +
 \frac{ l (l+1)}{x^2}
\end{equation}

 First, we estimate our
approximation in the case of the harmonic oscillator potential $
V(r) = \alpha r^2 $ for which the exact wave functions
$\psi_{ex}(x)$ are well known \cite{flug}. Figures 1, 2 and 3 show that
the proposed approximation gives fairly accurate wave functions in
this case. Here  solid lines reproduce $\psi_{app}(x)$ and dashed
lines reproduce $10 (\psi_{app}(x) - \psi_{ex}(x))$. Note that
without factor 10 the difference $(\psi_{app}(x) - \psi_{ex}(x))$
is invisible in comparison with $\psi_{app}(x)$.

 We can calculate the expectation values
\begin{equation}
  e_{app} =  <\psi_{app}|\hat H|\psi_{app}>
\end{equation}
 and
\begin{equation}
 (e'_{app})^2 =  <\psi_{app}|\hat H^2|\psi_{app}>
\end{equation}
 with the help of the normalized approximate wave  functions
 ($ <\psi_{app}|\psi_{app}> =1$). It should be stressed that $
 (e_{app}))^2 \neq  (e'_{app})^2 $ when the wave functions are
 not exact. Now we define the relative discrepancy
\begin{equation}
 d=\frac{e_{app}}{e'_{app}} -1.
\end{equation}

 We also calculate  relative virial error
\begin{equation}
 v =\frac{ < \psi_{app}| -\frac{d^2}{d x^2} +  \frac{ l
(l+1)}{x^2}|\psi_{app}>}{ < \psi_{app}| \frac{1}{2}k
x^k|\psi_{app}>} -1
\end{equation}
 which is equal to zero for the exact solutions.
 Finally, we characterize our approximation by the usual relative energy
error
\begin{equation}
\delta e = \frac{e_{app}}{e_{ex}} - 1
\end{equation}
where $e_{ex}$ is the exact energy value.

\begin{figure}[h!]
\centering
\includegraphics{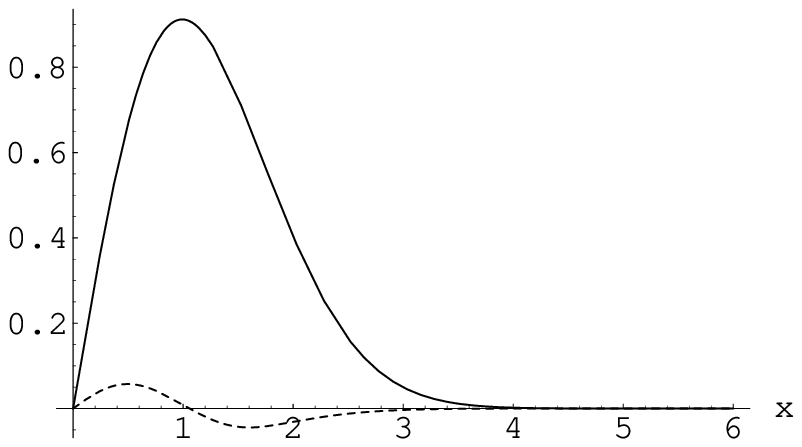} \caption{Harmonic oscillator radial wave function  \\for $l=0, n=0$.}
\end{figure}

\vspace{1cm}

\begin{figure}[h!]
\centering
\includegraphics{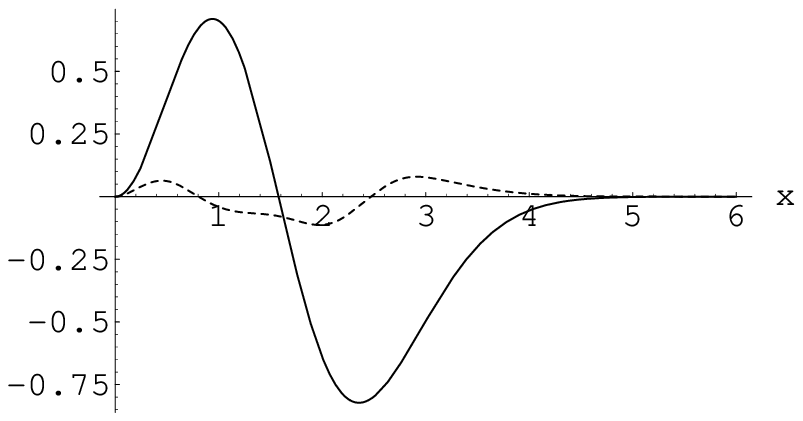}
\centering
\caption{Harmonic oscillator radial wave function  \\for $l=1, n=1$.}
\end{figure}

\vspace{1cm}

\begin{figure}[h!]
\centering
\includegraphics{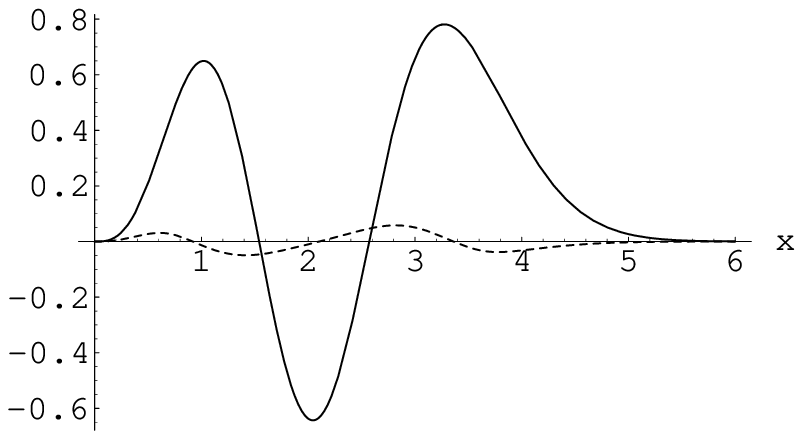} \caption{Harmonic oscillator radial wave function \\ for $l=2, n=2$.}
\end{figure}
\newpage

\begin{table*}
\caption{ Numerical verification of the proposed approximation in
the case of power-law potentials $V(r) =  \alpha r^k$ .}
\begin{tabular}{c| c| c| c| c| c| c}\hline
$l$ &$n$ &$v$ &$d$ &$e_{ex}$ &$\delta e$ &$\delta e$\cite{loba}
\\ \hline
 \multicolumn{6}{c}{$k=1$}  \\ \hline
 $0$&$0$&$0$&$0$&$2.33811$&$0$&$5.00 \cdot 10^{-3}$ \\
 $0$&$1$&$0$&$0$&$4.08795$&$0$&$1.50 \cdot 10^{-3}$ \\
 $0$&$2$&$0$&$0$&$5.52056$&$0$&$7.00 \cdot 10^{-4}$
 \\ \hline
 $1$&$0$&$2.841 \cdot 10^{-2}$&$-3.229 \cdot 10^{-3}$&$3.36126$&$1.020 \cdot 10^{-3}$&$1.20 \cdot 10^{-3}$ \\
 $1$&$1$&$1.269 \cdot 10^{-2}$&$-3.563 \cdot 10^{-5}$&$4.88445$&$3.890 \cdot 10^{-5}$&$7.00 \cdot 10^{-4}$ \\
 $1$&$2$&$8.420 \cdot 10^{-3}$&$-2.714 \cdot 10^{-5}$&$6.20762$&$2.739 \cdot 10^{-5}$&$4.00 \cdot 10^{-4}$
 \\ \hline
 $2$&$0$&$1.446 \cdot 10^{-2}$&$-1.607 \cdot 10^{-3}$&$4.24818$&$7.368 \cdot 10^{-4}$&$5.00 \cdot 10^{-4}$ \\
 $2$&$1$&$6.717 \cdot 10^{-3}$&$-7.023 \cdot 10^{-6}$&$5.62971$&$1.421 \cdot 10^{-5}$&$3.00 \cdot 10^{-4}$
 \\ \hline
 \multicolumn{6}{c}{$k=2$}  \\ \hline
 $0$&$0$&$1.588 \cdot 10^{-2}$&$-6.087 \cdot 10^{-5}$&$3$&$5.634 \cdot 10^{-5}$&-- \\
 $0$&$1$&$4.071 \cdot 10^{-3}$&$-7.104 \cdot 10^{-7}$&$7$&$1.889 \cdot 10^{-6}$&-- \\
 $0$&$2$&$1.939 \cdot 10^{-3}$&$-8.050 \cdot 10^{-8}$&$11$&$3.327 \cdot
 10^{-7}$&--
 \\ \hline
 $1$&$0$&$3.220 \cdot 10^{-2}$&$-9.040 \cdot 10^{-3}$&$5$&$1.766 \cdot 10^{-3}$&-- \\
 $1$&$1$&$1.174 \cdot 10^{-2}$&$-9.318 \cdot 10^{-5}$&$9$&$6.556 \cdot 10^{-5}$&-- \\
 $1$&$2$&$7.415 \cdot 10^{-3}$&$-5.584 \cdot 10^{-5}$&$13$&$4.615 \cdot
 10^{-5}$&--
 \\ \hline
 $2$&$0$&$1.614 \cdot 10^{-2}$&$-4.528 \cdot 10^{-3}$&$7$&$1.264 \cdot 10^{-3}$&--\\
 $2$&$1$&$5.545 \cdot 10^{-3}$&$-2.336 \cdot 10^{-5}$&$11$&$2.727 \cdot 10^{-5}$&-- \\
 $2$&$2$&$3.520 \cdot 10^{-3}$&$-7.132 \cdot 10^{-6}$&$15$&$6.667 \cdot
 10^{-6}$&--
 \\ \hline
 \multicolumn{6}{c}{$k=4$}  \\ \hline
 $0$&$0$&$2.931 \cdot 10^{-2}$&$-8.408 \cdot 10^{-4}$&$3.79967$&$2.543 \cdot 10^{-4}$&$-2.85 \cdot 10^{-2}$ \\
 $0$&$1$&$3.420 \cdot 10^{-3}$&$-1.627 \cdot 10^{-5}$&$11.6448$&$9.918 \cdot 10^{-6}$&$-5.30 \cdot 10^{-3}$ \\
 $0$&$2$&$7.311 \cdot 10^{-4}$&$-1.982 \cdot 10^{-6}$&$21.2384$&$1.963 \cdot 10^{-6}$&$-1.80 \cdot 10^{-3}$
 \\ \hline
 $1$&$0$&$4.215 \cdot 10^{-2}$&$-2.396 \cdot 10^{-2}$&$7.10845$&$3.011 \cdot 10^{-3}$&$-1.00 \cdot 10^{-2}$ \\
 $1$&$1$&$1.021 \cdot 10^{-2}$&$-2.418 \cdot 10^{-4}$&$16.0327$&$1.185 \cdot 10^{-4}$&$-3.80 \cdot 10^{-3}$ \\
 $1$&$2$&$5.524 \cdot 10^{-3}$&$-9.617 \cdot 10^{-5}$&$26.3500$&$6.831 \cdot 10^{-5}$&$-1.60 \cdot 10^{-3}$
 \\ \hline
 $2$&$0$&$2.090 \cdot 10^{-2}$&$-1.182 \cdot 10^{-2}$&$10.8424$&$2.103 \cdot 10^{-3}$&$-4.70 \cdot 10^{-3}$ \\
 $2$&$1$&$3.665 \cdot 10^{-3}$&$-8.348 \cdot 10^{-5}$&$20.6435$&$5.329 \cdot 10^{-5}$&$-3.00 \cdot 10^{-3}$ \\
 $2$&$2$&$1.630 \cdot 10^{-3}$&$-1.928 \cdot 10^{-5}$&$31.6147$&$1.265 \cdot 10^{-5}$&$-1.70 \cdot 10^{-3}$
 \\ \hline
\end{tabular}
\end{table*}
We calculate the values $\delta e$  for  the linear and quartic
potentials with respect to   the accurate numerical eigenenergies
\cite{popov}. We compare our results  with the recent results in
the framework of an integral semiclassical method for calculating
the spectra for spherically symmetric potentials \cite{loba}. In
the case of linear potential with $l=0$ our approach gives exact
eigenfunctions and eigenenergies. Table 1 demonstrates validity of
our approximation in the linear, quadratic and quartic cases .

\section{Conclusion}

An old problem in semiclassical analysis is the development of
global uniform approximations for the wave functions. In the
present paper, this problem has been solved by means of the
reconstruction of the WKB series and subsequent explicit summation
of the leading constituent (partial) series. Such approach  yields
satisfactory  description of the wave functions in the
important case of the radial Schr\"odinger equation with the
power-law potentials.

\end{document}